

%
\newread\epsffilein    
\newif\ifepsffileok    
\newif\ifepsfbbfound   
\newif\ifepsfverbose   
\newif\ifepsfdraft     
\newdimen\epsfxsize    
\newdimen\epsfysize    
\newdimen\epsftsize    
\newdimen\epsfrsize    
\newdimen\epsftmp      
\newdimen\pspoints     
\pspoints=1bp          
\epsfxsize=0pt         
\epsfysize=0pt         
\def\epsfbox#1{\global\def\epsfllx{72}\global\def\epsflly{72}%
   \global\def\epsfurx{540}\global\def\epsfury{720}%
   \def\lbracket{[}\def\testit{#1}\ifx\testit\lbracket
   \let\next=\epsfgetlitbb\else\let\next=\epsfnormal\fi\next{#1}}%
\def\epsfgetlitbb#1#2 #3 #4 #5]#6{\epsfgrab #2 #3 #4 #5 .\\%
   \epsfsetgraph{#6}}%
\def\epsfnormal#1{\epsfgetbb{#1}\epsfsetgraph{#1}}%
\def\epsfgetbb#1{%
%
%
\openin\epsffilein=#1
\ifeof\epsffilein\errmessage{I couldn't open #1, will ignore it}\else
%
%
   {\epsffileoktrue \chardef\other=12
    \def\do##1{\catcode`##1=\other}\dospecials \catcode`\ =10
    \loop
       \read\epsffilein to \epsffileline
       \ifeof\epsffilein\epsffileokfalse\else
%
%
          \expandafter\epsfaux\epsffileline:. \\%
       \fi
   \ifepsffileok\repeat
   \ifepsfbbfound\else
    \ifepsfverbose\message{No bounding box comment in #1; using defaults}\fi\fi
   }\closein\epsffilein\fi}%
%
%
%
\def\epsfclipoff{\def\epsfclipstring{\ifepsfdraft\space clip\fi}}%
\epsfclipoff
\def\epsfsetgraph#1{%
   \epsfrsize=\epsfury\pspoints
   \advance\epsfrsize by-\epsflly\pspoints
   \epsftsize=\epsfurx\pspoints
   \advance\epsftsize by-\epsfllx\pspoints
%
%
   \epsfxsize\epsfsize\epsftsize\epsfrsize
   \ifnum\epsfxsize=0 \ifnum\epsfysize=0
      \epsfxsize=\epsftsize \epsfysize=\epsfrsize
      \epsfrsize=0pt
%
%
     \else\epsftmp=\epsftsize \divide\epsftmp\epsfrsize
       \epsfxsize=\epsfysize \multiply\epsfxsize\epsftmp
       \multiply\epsftmp\epsfrsize \advance\epsftsize-\epsftmp
       \epsftmp=\epsfysize
       \loop \advance\epsftsize\epsftsize \divide\epsftmp 2
       \ifnum\epsftmp>0
          \ifnum\epsftsize<\epsfrsize\else
             \advance\epsftsize-\epsfrsize \advance\epsfxsize\epsftmp \fi
       \repeat
       \epsfrsize=0pt
     \fi
   \else \ifnum\epsfysize=0
     \epsftmp=\epsfrsize \divide\epsftmp\epsftsize
     \epsfysize=\epsfxsize \multiply\epsfysize\epsftmp   
     \multiply\epsftmp\epsftsize \advance\epsfrsize-\epsftmp
     \epsftmp=\epsfxsize
     \loop \advance\epsfrsize\epsfrsize \divide\epsftmp 2
     \ifnum\epsftmp>0
        \ifnum\epsfrsize<\epsftsize\else
           \advance\epsfrsize-\epsftsize \advance\epsfysize\epsftmp \fi
     \repeat
     \epsfrsize=0pt
    \else
     \epsfrsize=\epsfysize
    \fi
   \fi
%
%
   \ifepsfverbose\message{#1: width=\the\epsfxsize, height=\the\epsfysize}\fi
   \epsftmp=10\epsfxsize \divide\epsftmp\pspoints
   \vbox to\epsfysize{\vfil\hbox to\epsfxsize{%
      \ifnum\epsfrsize=0\relax
        \includegraphics{\ifepsfdraft}%
      \else
        \epsfrsize=10\epsfysize \divide\epsfrsize\pspoints
        \includegraphics{\ifepsfdraft}%
      \fi
      \hfil}}%
\global\epsfxsize=0pt\global\epsfysize=0pt}%
%
%
{\catcode`\%=12 \global\let\epsfpercent=
%
%
\long\def\epsfaux#1#2:#3\\{\ifx#1\epsfpercent
   \def\testit{#2}\ifx\testit\epsfbblit
      \epsfgrab #3 . . . \\%
      \epsffileokfalse
      \global\epsfbbfoundtrue
   \fi\else\ifx#1\par\else\epsffileokfalse\fi\fi}%
%
%
\def\epsfempty{}%
\def\epsfgrab #1 #2 #3 #4 #5\\{%
\global\def\epsfllx{#1}\ifx\epsfllx\epsfempty
      \epsfgrab #2 #3 #4 #5 .\\\else
   \global\def\epsflly{#2}%
   \global\def\epsfurx{#3}\global\def\epsfury{#4}\fi}%
%
%
\def\epsfsize#1#2{\epsfxsize}
%
%


\newcount\xrefpos \xrefpos=0
\newcount\yrefpos \yrefpos=0
\newcount\xput \xput=0
\newcount\yput \yput=0
\def\refpos#1 #2 #3{\global\xrefpos=#1 \global\yrefpos=#2
                         \rlap{$\smash{#3}$}}
\def\put #1 #2 #3{\xput=#1 \yput=#2
                  \advance\xput by -\xrefpos
                  \advance\yput by -\yrefpos
                  \rlap{\kern\the\xput truebp
                        \vbox to 0pt{\vss\hbox{$\displaystyle #3$}%
                        \kern\the\yput truebp}}}
\def\beginlabels\refpos#1\endlabels{\hbox{$\refpos#1$}}

\font\mybb=msbm10 at 12pt
\def\bb#1{\hbox{\mybb#1}}
\def\bZ {\bb{Z}}
\def\bR {\bb{R}}

\def\bN {\bb{N}}



\tolerance=10000
\input phyzzx
\input epsf.tex
 \def\unit{\hbox to 3.3pt{\hskip1.3pt \vrule height 7pt width 
.4pt \hskip.7pt
\vrule height 7.85pt width .4pt \kern-2.4pt
\hrulefill \kern-3pt
\raise 4pt\hbox{\char'40}}}

\def\arc {\rm{arc}}
\REF\bog {E.B. Bogomol'nyi, Sov. J. Nucl. Phys. {\bf 24} (1976) 449.}
\REF\wo {E. Witten and D. Olive, Phys. Lett. {\bf B78} (1978) 97.}
\REF\agf{L.Alvarez-Gaum\'e and D.Z. Freedman, Commun. Math. Phys. {\bf
91} (1983) 87.}
\REF\hpt{C.M. Hull, G. Papadopoulos and P.K. Townsend, Phys. Lett {\bf
B316} (1993) 291.}
\REF\gpta{G. Papadopoulos and P.K. Townsend, Class. Quantum Grav. {\bf
11} (1994) 515; {\bf 11} (1994) 2163.}
\REF\pt{E.R.C. Abraham and P.K. Townsend,  Phys. Lett. {\bf B291}
(1992) 85; {\bf B295} (1992) 225.}
\REF\gptb{G. Papadopoulos and P.K. Townsend, Nucl. Phys. {\bf B444}
(1995) 245.}
\REF\hps{P.S. Howe, G. Papadopoulos and K.S. Stelle, Nucl. Phys. {\bf
B296} (1988) 26.}
\REF\manton{N. Manton, Phys. Lett. {\bf 110B} (1982) 319. }
\REF\athit{M. Atiyah and N. Hitchin, {\it The geometry and dynamics of
magnetic monopoles}, Princeton
University press (1988), New Jersey.}
\REF\Dynkin{E.B. Dynkin, American Math. Soc. Trans. Series 2 {\bf 6},
 (1957) 111.}
\REF\raj{R. Rajaraman, {\it An introduction to solitons and instantons
in quantum field theory},
North-Holland, Amsterdam (1989).}
\REF\witten{E. Witten, Commun. Math. Phys. {\bf 92 } (1984), 455 }


\Pubnum{ \vbox{ \hbox{R/96/45} } }
\pubtype{}
\date{October 1996}

\titlepage

\title {\bf Solitons in (1,1)-supersymmetric massive sigma models}

\author{A. Opfermann and G. Papadopoulos}
\address{DAMTP, Silver St., 
\break
Cambridge CB3 9EW, U.K.}

\abstract{We find the solitons of massive (1,1)-supersymmetric sigma models
with target space the groups $SO(2)$ and $SU(2)$ for a class of scalar
potentials and compute their charge,
mass and moduli space metric.  We also investigate the massive sigma models
with target space any semisimple Lie group and show that some of their
solitons can be obtained from
embedding the $SO(2)$ and $SU(2)$ solitons.}

\endpage


\chapter{Introduction}

It has been known for some time that some of the non-perturbative aspects of
field theories can be understood by examining the properties of their 
soliton solutions. 
This has recently been emphasized both in
four-dimensional Yang-Mills theory and in string theory. Typically,
field theories with solitons are those for
which a lower bound can be established for the energy  in terms of the values
of their Noether and topological charges. The solitons are the
classical solutions that saturate
this bound. Therefore the solitons are also solutions of the
Bogomol'nyi equations which enforce the saturation
of the bound [\bog]. The Bogomol'nyi equations are usually first order
differential equations in spacetime
derivatives and simpler to solve than the  field equations of a theory
which are quadratic.
Although a bound for the energy can be established for a large class 
of theories, it is
better understood in the context of the supersymmetric ones.  In this
case, the bound is a consequence
of the supersymmetry algebra, and the topological and the Noether
charges that appear in the bound are central
charges of this algebra [\wo]. 

Examples of such theories are the  two-dimensional
(p,q)-supersymmetric massive sigma models, with
or without torsion [\agf,\hpt,\gpta]. Apart from the usual metric and 
torsion couplings of sigma models, the
bosonic part of the action of these theories has a scalar potential.
The energy is bounded from below by the
value of various topological and Noether charges all of which are
central charges of the (p,q)-supersymmetry
algebra. Most of these models have solitons\foot{We use the term
`soliton' for every solution of the
Bogomol'nyi equations that has finite energy.} that saturate the bound
and interpolate between the different classical vacua; some explicit
soliton solutions have been
found in [\pt,\gptb].  The class of massive sigma models with torsion 
includes the
$G$-models, {\rm i.e.} the  (1,1)-supersymmetric massive sigma models
for which the target space is a group manifold $G$
[\gptb]. The bosonic part of the action of these models consists of
the action of a  Wess-Zumino-Witten model
with target space a Lie group
$G$ and the scalar potential  
$$
V={m^2\over4}g^{ij}\big(2\kappa_A L^A_i+\partial_ia\big) \big(2\kappa_B\,
L^B_j+\partial_ja\big)\ ,
\eqn\onea
$$
where $g$ is the bi-invariant metric on $G$, $L$ is the left invariant
frame, $m$ is a constant with
dimension that of a mass,
$\kappa$ is a constant vector in the Lie algebra, ${\cal L}(G)$, of
$G$,  $A=1,...,\rm{dim}\,{\cal L}(G)$,
and
$$
a={1\over n}{\rm tr}\,k^n\ ,
\eqn\oneaa
$$
$n\in \bZ$, $n\not=0$, and $k\in G$. Note that the scalar potential
\onea\  is specified by two parameters,
the coupling constant $\kappa$ and the integer $n$. Both the vacuum
structure of these models and the
properties of their solitons depend on the values of these two parameters.

In this paper, we shall systematically investigate the solutions of
the  Bogomol'nyi equations of $G$-models for which the target space is
any  semisimple compact Lie group $G$. It is  sufficient to
investigate those models for which $n>0$ because the rest are related
to the $n\geq 0$ ones by a field
redefinition. It can also be arranged, without loss of generality, for the
coupling constant
$\kappa$ to lie in the Cartan subalgebra of the group $G$.  With
this choice for $\kappa$, we shall show that all vacua lie on a
maximal torus of the group $G$.   We shall
first solve the Bogomol'nyi equations of $SO(2)$-models and show that
the $SO(2)$ solitons generalise the sine-Gordon solitons; the latter
are recovered in
the limit $\kappa=0$. We shall then give the soliton solutions of the
$SU(2)$ model for $n=1$ and $n=2$ without
posing any restriction on the coupling constant $\kappa$.  In the
$n=1$ case, we shall find two classes of
solitons.  The first one is a set of static solitons that lie on a
maximal torus of $SU(2)$
and are embeddings of the $SO(2)$ solitons.  The second one is a set
of time-dependent solutions that, apart
from their asymptotic values, lie in the complement of the maximal
torus of $SU(2)$ and include the solution
given in [\gptb]. In the
$n=2$ case, we shall show that all solitons which lie on the maximal  
torus are obtained from embedding
of $SO(2)$ solitons into the $SU(2)$ model. 
In addition, we shall examine the qualitative properties of the Bogomol'nyi
equations for $SU(2)$  $n>2$ models and support our conclusions with a
numerical calculation for the $n=3, 4,
5$ cases (Fig [1,2,3]).
Finally, we shall show that the $G$-model for
$G$ any semisimple Lie group has static and time-dependent solutions.
The former lie on the maximal torus
of $G$ while the latter lie on the complement of the maximal torus.
Some of the $G$-model solitons 
are obtained from embedding $SO(2)$ and $SU(2)$ solitons.

The organisation of the paper is as follows:  In section two, we shall
introduce the Bogomol'nyi equations of
$G$-models and discuss their general properties. In section
three,  we shall solve the Bogomol'nyi equations for $G=SO(2)$ and
compute the charge, mass and moduli space
metric of the solitons. In section four, we shall examine the general
properties of the $SU(2)$ model
and give all static solitons that lie on the maximal torus of $SU(2)$.
In section five, we shall investigate
the time-dependent solitons  of the
$SU(2)$ $n=1$ model and give their charge, mass and moduli space
metric.  In section six, we shall
examine the solitons of the $SU(2)$ $n=2$ model and find that all of
them are static and lie on the maximal
torus. In section seven, we shall study the qualitative properties of
the Bogomol'nyi equations for the
$SU(2)$ $n>2$ models. The solitons for
models with any semisimple group $G$ as target space will be
investigated in section eight, and in section nine we
shall give our conclusions and comment on the quantum mechanical
properties of the solitons of $G$-models.


\chapter { The Bogomol'nyi equations}

Let ${\cal N}$ be a Riemannian manifold with metric $g$, a locally
defined two form $b$ and a Killing vector
$X$. The bosonic part of the action of (1,1)-supersymmetric massive
sigma models with torsion $H$ and scalar
potential
$V$ [\gptb]  is
$$
I= \int d^2x \big[ (g+b)_{ij} \partial_+ \phi^i \partial_-  
\phi^j-V(\phi)\big]\ ,
\eqn\onebb
$$
where $\phi$ is a map from the two-dimensional Minkowski spacetime,
with light-cone co-ordinates
$(x^+, x^-)$, into ${\cal N}$ and the two-form $b$ is the `gauge
potential' of the
torsion three-form $H$, $H_{ijk}={3\over2}\partial_{[i} b_{jk]}$.
Furthermore, the scalar potential is
$$
V={m^2\over4} g_{ij} \big( X^i X^j+  u^i u^j\big)\ ,
\eqn\potone
$$
where the one-form $u$ is orthogonal to the Killing vector $X$ ($X^i
u_i=0$) and
$$
X^i H_{ijk}=\partial_{[j} u_{k]}\ .
\eqn\pottwo
$$
It is clear that the supersymmetric vacua of the theory are the points
of the target space ${\cal N}$ where
both
$X$ and $u$ vanish. The action of the (1,1)-supersymmetric massive
sigma model is invariant under the
transformations
$$
\delta\phi^i=\eta X^i(\phi)
\eqn\ptthree
$$
with associated charge 
$$
Q=\int \, dx\, \big(X_i\,\partial_t\phi^i+u_i\, \partial_x\phi^i\big)\ ,
\eqn\potfour
$$
 where $\eta$ is an infinitesimal parameter and  $(t,x)$ are the
co-ordinates of two-dimensional Minkowski spacetime. 
The  charge $Q$ is the charge that appears as central charge in the
(1,1)-supersymmetry
algebra [\gptb].  The energy of the model is
$$
E={1\over2} \int\, dx\, \big(g_{ij}\, \partial_t\phi^i\,  
\partial_t\phi^j+g_{ij}\, \partial_x\phi^i\,
\partial_x\phi^j+V(\phi)\big)\ ,
\eqn\fivebb
$$
and the  bound of $E$ in terms of $Q$ is
$$
E\geq {m\over2} |Q|\ .
\eqn\sixbb
$$
We can split the charge $Q$ into a Noether charge $Q_N$ and a  
topological charge $Q_T$, $Q=Q_N+Q_T$. Although
there is no natural way to split $Q$, we can use the Hodge decomposion
of $u$ as a sum of an exact form
$\alpha$, a co-exact form
$\beta$ and a harmonic form
$\gamma$, ($u=\alpha+\beta+\gamma$), with respect to the metric $g$ to define
$$
Q_T=\int(\alpha+\gamma)_i\partial_x\phi^i\ .
\eqn\potsix
$$
and $Q_N\equiv Q-Q_T$;  $Q_N$ and $Q_T$ are separately
conserved. Both the fundamental states and the solitons of the theory
are charged with respect to the Noether
charge $Q_N$ but only the latter are charged with respect to the    
topological charge $Q_T$ as well.

In the special case of $G$-models with scalar potential \onea, we
choose the couplings as follows:
$g$ is the bi-invariant metric 
$$
g_{ij}=\delta_{AB}L^A{}_i L^B{}_j=\delta_{AB}R^A{}_i R^B{}_j
\eqn\metrone
$$
on $G$,
$H$ is the bi-invariant closed three form 
$$
H_{ijk}=-{1\over2}f_{ABC}L^A_iL^B_jL^C_k=-{1\over2} f_{ABC} R^A_iR^B_jR^C_k
\eqn\twobb
$$
on $G$, $X$ is the vector field
$$
X^i=\kappa^A (L-R)^i_A\ ,
\eqn\potseven
$$
and the one-form $u$ is
$$
u_i=\kappa_A (L+R)_i^A+\partial_ia\ ,
\eqn\poteight
$$
where $L$ is the left invariant frame of $G$, $R$ is the right
invariant frame of $G$, $f$ are the
structure constants of ${\cal L}(G)$, $A,B,C=1,...,{\rm dim}\,{\cal L}(G)$ and 
$i,j,k=1,...,{\rm dim}\, G$, $\kappa$ is a vector in the Lie algebra
of $G$ as in \onea\ and $a$ is given
in \oneaa.

The supersymmetric classical vacua of the action \onebb\ are the
solutions of the algebraic equation
$$
2\kappa_A+{\rm tr}\big( k^n t_A\big)=0
\eqn\fourb
$$
for $k\in G$,
where $\{t_A; A=1,...,{\rm dim}\,{\cal L}(G)\}$ is a basis of the Lie
algebra of the group $G$.
The charge $Q$ \potfour\ of $G$-models is
$$
Q=\int dx \big(\kappa_A (L-R)^A_i\partial_t\phi^i+\kappa_A  
(L+R)^A_i\partial_x\phi^i +\partial_x a \big)\ .
\eqn\fourbb
$$
For $G$ abelian, $L=R$ and $\kappa_A(L+R)^A$ is a closed one-form, so
$Q_N=0$ and
$$
Q_T=Q=\int \, dx\, \big(2\kappa_A L^A_i\partial_x\phi^i +\partial_x a \big)\ .
\eqn\potninea
$$  
However, for $G$ semisimple, the one-form $u-da$ is  co-exact and the
topological charge $Q_T$ \potsix\  is
simply
$$
Q_T=\int dx \partial_x a\ .
\eqn\potnine
$$

To find the mass and charge of the bosonic fundamental  states of
$G$-models, we use the background field method (see for example
[\hps]) to linearise the theory around a
classical supersymmetric vacuum $k_0$. The mass-square and charge    
matrices written in the left-invariant
frame basis are then
$$
{\cal M}_{AB}={m^2\over4}\big(\kappa^C \kappa^D f^E{}_{AC}\, f_{EBD}+
n^2 \sum_C{\rm tr}\,(t_{(A}\, t_{C)}
k_0^n)\,{\rm tr}(t_{(B}\, t_{C)} k_0^n)\big)\ ,
\eqn\potten
$$
and
$$
{\cal Q}_{AB}=-\kappa^C f_{ABC}\ ,
\eqn\poteleven
$$
respectively. The eigenvalues of  ${\cal M}$ are the  square of the 
masses of the fundamental
states and the eigenvalues of ${\cal Q}$ are their charges.  To do
this computation, we have used
\pottwo\ to show that
$du=0$ evaluated at any supersymmetric vacuum.  The fundamental states
parallel to the zero modes of the charge-matrix
\poteleven\ have zero Noether charge. The zero modes of ${\cal Q}$
are orthogonal to the tangent vectors of
the coadjoint orbit of $G$ determined by $\kappa$ and therefore
the directions that carry the charge are those
along the coadjoint orbit; in fact ${\cal Q}$ is the symplectic form
of the coadjoint orbit. The mass of the
fundamental states need not saturate the bound \sixbb\ since
 $$
{\cal M}\geq {m^2\over 4} {\cal Q} {\cal Q}^t
\eqn\potelevenb
$$
due to the presence of the second term in the mass-square-matrix
\potten. So the fundamental states obey the
bound as expected for supersymmetric theories but they are not  
necessarily BPS states.  Apart from the bosonic
fundamental states, the (1,1)-supersymmetric sigma models have also 
fermionic ones which can be investigated
in a similar way; for the purpose of this paper we shall restrict our 
attention  to the bosonic fundamental states.

The classical configurations that saturate the bound \sixbb\ satisfy
the following Bogomol'nyi type equations:
$$
\eqalign{
\partial_t k(x,t)&=\mp {m\over2} (\kappa k-k \kappa)
\cr
\partial_x k(x,t)&=\mp {m\over2} \big(\kappa k+k \kappa +k \sum_A t_A
{\rm tr}(t_A k^n)\big)\ .}
\eqn\oneb
$$
The first of these equations can be easily solved by setting
$$
k(x,t)=\exp\big(\mp {m\over2} \kappa t\big) k(x) \exp\big(\pm
{m\over2} \kappa t\big)\ .
\eqn\twob
$$
Substituting this into the second equation of \oneb, it reduces to 
$$
{d \over dx}k(x)=\mp {m\over2} \big(\kappa k(x)+k(x) \kappa +k(x)
\sum_A t_A {\rm tr}(t_A
k^n(x))\big)\ .
\eqn\threeb 
$$
This is a non-linear but ordinary differential equation and the 
investigation of its solutions for various
choices of $G$ and $n$ will be our task in the rest of the paper. 

It is sufficient to investigate \threeb\ for
$n>0$ because the transformation
$$
k\rightarrow k^{-1}, \qquad x\rightarrow -x 
\eqn\fiveba
$$
acting on a model with coupling constants $g, b,\kappa, m$  and
$a={1\over n}{\rm tr}\, k^n$ transforms it to a
model with coupling constants $g, b,\kappa, m$ and
$a=-{1\over n}\,{\rm tr}\, k^{-n}$.  Therefore the solitons of models 
with $n<0$  can be obtained from
the solitons of models with $n>0$ by acting on the latter with \fiveba.  
In addition, the Lagrangian \onebb, the field equations and the
Bogomol'nyi equations of the
(1,1)-supersymmetric sigma model are invariant under the sigma model symmetry
$$
k(x,t)\rightarrow h k(x,t) h^{-1} \qquad \kappa\rightarrow h \kappa h^{-1}\ ,
\eqn\fiveb
$$
 where $h\in G$.  Although sigma
model symmetries are not associated with conserved charges, we can use
\fiveb\ to choose, without loss of
generality, the coupling constant $\kappa$ in the Cartan subalgebra of
${\cal L} (G)$. This choice
of $\kappa$ simplifies both the Bogomol'nyi equations \oneb\ and the 
equation for the vacua  \fourb\ of the
model. In particular, we shall show in chapter eight that it can be
arranged for the vacua to lie in the
maximal torus of the group $G$.

The solutions of the Bogomol'nyi equations may depend on several
moduli parameters.  In that case, as
for BPS monopoles [\manton], one can define a metric on the moduli
space of solutions as
follows:
$$
ds^2={1\over2}\int dx\, g_{ij} d\phi^i d\phi^j\ ,
\eqn\sevenb
$$
where the differential on $\phi$ denotes variation with respect to the
moduli parameters. The soliton
solutions of the Bogomol'nyi equations \oneb\ preserve $1/2$ of the 
supersymmetry and therefore their
effective theory is an $N=1$ supersymmetric one-dimensional sigma
model with target space the moduli space and
metric the moduli metric \sevenb.

\chapter{The SO(2) model}

The simplest of the $G$-models is the one with the group $SO(2)$ as
target space.
Since the target space is one dimensional, the antisymmetric tensor
coupling $b$ is identically zero.  A
parameterisation of the group manifold
$SO(2)$ is 
$$
k=\pmatrix{\cos\theta & \quad -\sin\theta\cr \sin\theta & \quad \cos\theta}
\eqn\onec
$$
and a basis in ${\cal L}\big(SO(2)\big)$ is
$$
t_1=\pmatrix{0&\quad -1\cr 1&\quad 0}\ ,
\eqn\onecnew
$$
where $0\leq\theta<2\pi$.
To find the classical vacua of the $SO(2)$ model, we use the same
symbol $\kappa$ for the vector $\kappa$ and
its component along $t_1$ and
 substitute \onec\ into the equation for the vacua \fourb\ to find
$$
\sin (n \theta)=\kappa\ .
\eqn\twoc
$$
For $|\kappa|>1$, the model does not have any supersymmetric vacua and
therefore supersymmetry is spontaneously
broken. For $0\leq \kappa\leq 1$, the supersymmetric vacua of the
$SO(2)$ model are
$$
\theta[\tilde \ell, \pm]={1\over n}\big(\pm\theta_0+ {1\mp 1\over 2} 
\pi + 2\pi \tilde\ell\big) \ , 
\eqn\threec
$$
where $\tilde\ell$ is an integer,  $0\leq \tilde\ell< n$, and
$\sin\theta_0=\kappa$,  $0\leq \theta_0<{\pi\over2}$; the angle 
$\theta_0$ vanishes for $\kappa=0$. This
$2n$ vacua can be ordered by the size of the angle
$\theta$ as follows:
$$
\theta[0,+]<\cdots<\theta[\tilde\ell,+]<\theta[\tilde\ell,-]<
\theta[\tilde\ell+1,+]<\cdots<\theta[n-1,-]\ .
\eqn\threeorder
$$
For $\kappa=1$, $\theta_0={\pi\over2}$,  there are only $n$ 
supersymmetric vacua (without counting multiplicities) and the
equation for the vacua \threec\ can be 
rewritten as
$$
\theta[\tilde \ell]={\pi\over n} ({1\over2}+2 \tilde \ell)\ .
\eqn\fourcc
$$
 We can order these vacua with respect to the value of $\tilde\ell$.  
We shall not investigate the models with $-1\leq\kappa<0$ since it is 
easy to generalise the results obtained
in the models with $0\leq \kappa\leq 1$ to this case.

All  $SO(2)$ solitons are static. To find them, we must solve
\threeb\ which rewritten in the co-ordinates
\onec\ becomes
$$
{d\over dx} \theta=\pm m \big(-\kappa+\sin(n \theta)\big)\ .
\eqn\fourc
$$
We shall consider the following three cases: (i)
For $0<\kappa<1$, a linearisation of the equation \fourc\ with the
plus sign around
the vacua $\theta[\tilde \ell,\pm]$  reveals that the vacua 
$\theta[\tilde \ell, +]$ are
`sources' while the vacua  $\theta[\tilde\ell, -]$  are `sinks'. 
(If we choose the minus sign in
\fourc\ then the sources and sinks will be interchanged).  This
suggests that there should be
solutions that interpolate between neighbouring sources and sinks.  
Indeed the solutions that
interpolate between the vacua $\theta[\ell, +]$ and $\theta[\ell, -]$ are 
$$
\theta(x)={2\over n} \bigg( {\arc}\tan\bigg[ {1\over \kappa}
\big[1+\sqrt{1-\kappa^2} {\tanh}\big(\pm 
{n m\over2}\sqrt{1-\kappa^2} (x-x_0)\big)\big]\bigg]+\ell \pi\bigg)\ ,
\eqn\fivec
$$
where $0\leq \ell< n$, $\ell\in \bN$, and $x_0$ is an integration
constant due to the
translation invariance of the underlying model. 
In addition, the  solutions
$$
\theta(x)=\cases{
{2\over n}
  \bigg({\arc}\tan\bigg[{1\over \kappa}\big[1+\sqrt{1-\kappa^2}
{\coth}\big(\pm  {n m\over2}\sqrt{1-\kappa^2} (x-x_0)\big)\big]\bigg]
\cr \qquad 
+(\ell+1)
\pi\bigg)\ ,\qquad\qquad\qquad\pm x<0
\cr
{2\over n}\bigg({\arc}\tan\bigg[ {1\over \kappa}\big[1+\sqrt{1-\kappa^2}
{\coth}\big(\pm {n m\over2}\sqrt{1-\kappa^2} (x-x_0)\big)\big]\bigg]
\cr \qquad
+\ell
\pi\bigg)\ , \qquad\qquad\qquad \qquad \pm x>0 }
\eqn\fivecc
$$
interpolate between the vacua $\theta[\ell, -]$ and
$\theta[ \ell+1, +]$, for $0\leq\ell<n-1$, and the  vacua  
$\theta[n-1, -]$ and $\theta[0, +]$
for $\ell=n-1$. We remark that the solutions \fivecc\ are continuous 
and differentiable at $x=0$. The solitons
are those solutions in \fivec\ and in \fivecc\ with the plus sign 
whereas the anti-solitons are those with the
minus sign.  We note that the
$SO(2)$ solitons of \fivec\ and \fivecc\ interpolate between all
neighbouring pairs of vacua of the model.   (ii) For
$\kappa=0$, the equation \fourc\ becomes
 that of the sine-Gordon theory. The solutions are 
$$
\theta=\zeta\, {2\over n}\bigg({\rm arc}\tan\big[\exp\big(\pm m n 
(x-x_0)\big)\big]+\ell \pi\bigg)
\eqn\sixc
$$
for $0\leq \ell< n$, $\ell\in \bZ$ and $\zeta=\pm$.   (iii) For
$\kappa=1$ and $n>1$ , the solutions that interpolate between the
vacua $\theta[\ell]$ and $\theta[\ell+1]$,
for $0\leq\ell<n-1$, and the  vacua  $\theta[n-1]$ and $\theta[0]$
for $\ell=n-1$, are
$$
\theta={2\over n}\bigg({\rm arc}\tan\big[\mp m n (x-x_0)\big]-
{\pi\over4}+\ell \pi\bigg)\ . 
\eqn\sixcc
$$
These are $2n$ solutions, $n$ solitons and $n$ anti-solitons, 
interpolating between all
neighbouring pairs of vacua as in the previous two cases.

For $SO(2)$ models, the charge $Q$ \fourbb\ is equal to the
topological charge $Q_T$ \potsix.  The value of $Q$
for the solitons is
$$
Q=\mp {4\over n} \bigg[\sqrt{1-\kappa^2}+\kappa\, {\rm arc}
\sin(\kappa)+\xi\, {\kappa\over2}\pi\bigg]
\ ,
\eqn\sevenc
$$
where $\xi$  is either minus or plus, ($\xi=\mp$);  $Q$ for $\xi=-$ 
is the charge of \fivec\ solutions and $Q$ for
$\xi=+$ is the charge of
\fivecc\ solutions. From this, it is straightforward to compute the
mass of the solitons to find\foot{We have
identified the energy with the mass of a soliton configuration.}
$$
E={2 m\over n}\bigg[\sqrt{1-\kappa^2}+\kappa\, {\rm
arc}\sin(\kappa)+\xi\, {\kappa\over2}\pi\bigg]\ .
\eqn\eightc
$$
The $SO(2)$ solitons \fivec\ and \fivecc\ have one moduli parameter
$x_0$. Using
\sevenb,  the metric on the moduli space is
$$
ds^2={ m\over n}\bigg[\sqrt{1-\kappa^2}+\kappa\, {\rm arc}
\sin(\kappa)+\xi\, {\kappa\over2}\pi\bigg]
dx_0^2={E\over 2} dx_0^2\ .
\eqn\ninec
$$
The charge, the mass and the moduli metric of the solitons \sixc\ 
are derived by setting $\kappa=0$ in the
above corresponding expressions.

Finally,  the charge, the mass and the moduli metric of the solutions 
\sixcc\  of models with $\kappa=1$ are 
$$
Q=\mp {4\over n} \pi\ ,
\eqn\tenc
$$
$$
E={2 m\over n} \pi\ ,
\eqn\tencc
$$
and
$$
ds^2={m\over n} \pi dx_0^2={E\over2} dx_0^2\ ,
\eqn\tenccc
$$
respectively. 
 
\chapter{The $SU(2)$ model}

A parameterisation of $SU(2)$ in terms of $2\times2$ matrices is
$$
k=\pmatrix{M+iN &\quad -U+iW \cr U+iW &\quad M-iN}; \quad M^2+N^2+U^2+W^2=1\ ,
\eqn\oned
$$
and a choice of basis in the Lie algebra ${\cal L}\big(SU(2)\big)$ of 
$SU(2)$ is
$$
t_1=i\pmatrix{1&\quad 0\cr 0&\quad -1}\ , \quad t_2=i
\pmatrix{0&\quad 1\cr 1&\quad 0}\ ,\quad
t_3=\pmatrix{0&\quad -1\cr 1&\quad 0}\ .
\eqn\twod
$$
Using the observation in section 2 that the Bogomol'nyi equations 
are invariant under the adjoint
action of $G$ on both the fields and the coupling constant $\kappa$, 
we can choose, without loss of generality,
$\kappa$ in the Cartan subalgebra of
${\cal L}\big(SU(2)\big)$, {\rm i.e.}
$$
\kappa=\kappa t_1\ .
\eqn\threed
$$
We have denoted both the vector $\kappa$ and its component along 
$t_1$ with the same symbol. The
distinction between the two will be clear from the context.

The vacua of the theory \fourb\  in the above  parameterisation are
the solutions of the equation 
$$
k^n=\pmatrix{\mp \sqrt{1-\kappa^2}+i\kappa &\quad 0\cr 0&\quad \mp 
\sqrt{1-\kappa^2}-i\kappa}\ ,
\eqn\fourdn
$$
in terms of $k$, $n\in \bN$.
It is clear that for $|\kappa|>1$, there are no supersymmetric vacua
and so supersymmetry is
spontaneously broken.  For $|\kappa|<1$ there are $2n$ supersymmetric 
vacua, and for $|\kappa|=1$
there are $n$ supersymmetric vacua (without counting multiplicities 
in the latter case). We shall restrict
our attention to $\kappa\geq0$ because it is straightforward to
generalise the results below to $SU(2)$
models with $\kappa<0$.  We remark that linearising the theory 
around the  vacua \fourdn, the
mass-square-matrix and charge-matrix of the fundamental states of 
the $SU(2)$ model are
$$
{\cal M}=m^2 \pmatrix{&1-\kappa^2\quad  &0\quad  &0\cr
				&0\quad &1\quad &0\cr
				&0\quad &0\quad &1}\ ,
\eqn\massmatone
$$
and 
$$
{\cal Q}=\pmatrix{&0\quad &0\quad &0\cr
				&0\quad &0\quad &-2\kappa\cr
				&0\quad &2 \kappa\quad &0}\ ,
\eqn\chargematone
$$
respectively.
Clearly for $|\kappa|<1$,  ${\cal M}>{m^2\over 4} {\cal Q} {\cal Q}^t$
and the fundamental states are not BPS
states. However, for
$|\kappa|=1$, ${\cal M}={m^2\over 4} {\cal Q} {\cal Q}^t$ and  all the
fundamental states become BPS with one
of them massless.

The equations \threeb\ in this parameterisation for the $SU(2)$ model are
$$
\eqalign{
{d\over dx}M\pm m [-\kappa N+(1-M^2) \tilde M_n]&=0
\cr
{d\over dx}N\pm m[\kappa M-NM\tilde M_n]&=0
\cr
{d\over dx}U\mp m U M \tilde M_n&=0
\cr
{d\over dx}W\mp m W M \tilde M_n&=0\ ,}
\eqn\fivedn
$$
where
$$
\tilde M_n=\sum^{[{n-1\over2}]}_{i=0}\pmatrix{n\cr 2i+1} 
(-1)^{i} M^{n-2i-1} (1-M^2)^{i}\ .
\eqn\sixdn
$$
There are three independent equations in \fivedn\ since the fourth one
is implied from the restriction
that $M, N, U, W$ lie on a 3-sphere.
Moreover from the last two equations of \fivedn, it is easy to see that 
$$
U=\lambda W\ ,
\eqn\sevendn
$$
where $\lambda$ is a real constant.  So we only need to determine
$M,N$ from the first two equations in
\fivedn; these equations  will be studied in the next three chapters 
for various values of $n$.

It is convenient for reasons that will become apparent later to 
parameterise $SU(2)$ in angular coordinates as follows:
$$
\eqalign{
M&=\cos\theta
\cr
N&=\sin\theta \cos\phi
\cr
U&=\sin\theta  \sin\phi \cos\psi
\cr 
W&=\sin\theta \sin\phi \sin\psi\ ,}
\eqn\nineda
$$ 
where $0\leq \theta, \phi<\pi$ and $0\leq \psi<2\pi$. In this 
parameterisation, the equation \fourdn\ for the
vacua can be written as
$$
\eqalign{
\sin(n\theta)&=\pm \kappa
\cr
\cos\phi&=\pm 1\ . }
\eqn\nfived
$$
Note that the vacua of the theory lie in $\phi=0$ and
$\phi=\pi$ semi-circle subspaces of
$SU(2)$, which are joined at $(\theta, \phi)=(0,0)$ and $(\theta, 
\phi)=(0,\pi)$, and at $(\theta,
\phi)=(\pi,0)$ and  $(\theta, \phi)=(\pi,\pi)$ to a circle in $SU(2)$;
in fact this circle is a maximal torus of
$SU(2)$. Solving \nfived\ for $0\leq \kappa <1$, we find that the $2n$
vacua of the model are
$$
\big(\theta, \phi\big)=\cases{\bigg({1\over
n}(\pm\theta_0+{1\mp1\over2}\pi+2\pi \tilde \ell), 0\bigg)\ ,
\qquad\qquad 0\leq\tilde
\ell\leq \big[{n-1\over2}\big]
\cr
\bigg({1\over n}\big(\pm\theta_0+{1\mp1\over2}\pi+2\pi (\tilde 
\ell+{1\over2})\big), \pi\bigg)\ ,\qquad 0\leq\tilde
\ell<\big[{n\over2}\big]\ ,}
\eqn\vacthree
$$ 
where $0\leq \theta_0<{\pi\over2}$ and $\sin \theta_0=\kappa$. 
For $\kappa=0$, $\theta_0$ vanishes and the
expression for the vacua becomes 
$$
\eqalign{
(\theta, \phi)&=\big({\ell \over n}\pi, 0\big)
\cr
(\theta, \phi)&=\big({\ell+1 \over n}\pi, \pi\big)\ , }
\eqn\qnew
$$
where $0\leq \ell<n$, $\ell \in \bN$. For $\kappa=1$, the $n$ vacua of
the theory (without counting
multiplicities) can be found by setting
$\theta_0=\pi/2$ in  \vacthree.

The equations \threeb\ written in the angular parameterisation are
$$
\eqalign{
{d\over dx}\theta&=\pm m[-\kappa \cos \phi+\sin(n\theta)]
\cr
{d\over dx}\phi&=\pm m\kappa \cot\theta \sin\phi
\cr
{d\over dx}\psi&=0\ .}
\eqn\ninedaan
$$
The third equation above implies that the angle $\psi$ is constant;  
$\psi$ is related to the ratio of
$U, W$ as $\cot\psi=\lambda$.  So it remains to solve the first two 
equations of \ninedaan\ for $\theta$ and
$\phi$ for the various values of $n$ and $\kappa$.

To find a class of solutions for the  equations \ninedaan\ above, we
set $\phi$ to its vacuum
values, {\rm i.e.}
$\phi=0,\pi$. This choice for $\phi$ solves the second of the
equations \ninedaan, so 
it remains to solve the first for $\theta$.
Since we have set $\phi=0,\pi$, the solutions lie on the
maximal torus of
$SU(2)$, and they are best described by extending the range of
$\theta$ from $0\leq
\theta<\pi$ in
\ninedaan\ to $0\leq \theta<2\pi$. Then, the first equation in
\ninedaan\ is precisely
the equation \fourc\ that we have encountered in the $SO(2)$
model. Moreover, the vacua 
\vacthree\ of the $SU(2)$ model can be identified with the vacua
\threec\ of the $SO(2)$ model as follows: 
$$
\theta[\ell,\pm]=\cases{\bigg({1\over
n}(\pm\theta_0+{1\mp1\over2}\pi+2\pi  \ell),
0\bigg)\ , \qquad\qquad 0\leq \ell\leq \big[{n-1\over2}\big]
\cr
\bigg({1\over n}\big(\pm\theta_0+{1\mp1\over2}\pi+2\pi (\tilde
\ell+{1\over2})\big), \pi\bigg)\ ,\qquad \big[{n-1\over2}\big]<
\ell<n \ ,}
\eqn\vacthreenew
$$
where $\tilde \ell=n-\ell-1$.
Furthermore,  as can
be easily seen from \twob, the solitons of the $SU(2)$ model that lie
on the maximal torus  are static.
From this we conclude that the solitons of the $SU(2)$ model that lie 
on the maximal torus are
embeddings of the solitons of the
$SO(2)$ model found in the previous section with 
$$ 
\theta\rightarrow \pmatrix{e^{i\theta}& \quad 0\cr 0& \quad e^{-i\theta}}\ ,
\eqn\fourteend
$$
for the associated value of $n$. The class of solutions that we have 
described includes those for which
$\kappa$ vanishes. This is because, if $\kappa=0$, the second equation
in \ninedaan\ implies that $\phi$ is
constant and  the asymptotic behaviour of the solitons requires that 
it should be set to its vacuum values. 

Note from \potninea\ and \potnine\ that the definition of the
topological charge of the $G$-models with
$G$ semisimple  is different from the definition of the topological
charge of the $SO(2)$ models.
The value of the topological charge $Q_T$ of any soliton of the
$SU(2)$ model, or any $G$-model with $G$
semisimple, interpolating between the two vacua
$(\theta,\phi)$ and $(\theta', \phi')$ is $Q_T=a(\theta',
\phi')-a(\theta, \phi)$ while for the $SO(2)$
solitons $Q_T=Q$. Therefore the value of the topological charge of the
$SO(2)$ solitons and the value of the
topological charge of their embeddings as $SU(2)$ solitons are
different.  Apart from the
$SU(2)$ solitons that lie on the maximal torus and we have examined
above, the $SU(2)$ model has other
solitons that interpolate between different vacua but otherwise lie 
entirely outside the maximal torus of
$SU(2)$. We shall investigate these solitons in the following three sections.

\chapter{The $SU(2)$ n=1 model}

As we have seen in the previous section, there are static solitons
that interpolate between the two
vacua of $SU(2)$ $n=1$ model with $0\leq \kappa<1$ and lie entirely on
the maximal torus of
$SU(2)$. Here, we shall find another class of solitons for this model 
that interpolate between
the two vacua but otherwise  lie  in the {\it complement} of the
maximal torus of $SU(2)$.
The equations \fivedn\ for $M,N$ in the $n=1$ case  are
$$
\eqalign{
{d\over dx}M\pm m [-\kappa N+(1-M^2)]&=0
\cr
{d\over dx}N\pm m[\kappa M-NM]&=0
\ .}
\eqn\fived
$$
Following [\gptb], we can find a class of solitons by setting 
$N=\kappa$. These solutions\foot{These
solutions are related to those of [\gptb] by a conjugation.}  are
$$
\eqalign{
M&=\sqrt{1-\kappa^2} \tanh\big(\mp m \sqrt{1-\kappa^2}(x-x_0)\big)
\cr
N&=\kappa
\cr
U&=\sqrt{1-\kappa^2} \cosh^{-1}\big( m \sqrt{1-\kappa^2}(x-x_0)\big) \cos\psi
\cr
W&=\sqrt{1-\kappa^2} \cosh^{-1}\big( m \sqrt{1-\kappa^2}(x-x_0)\big) \sin\psi
\ ,}
\eqn\eightd
$$
where $x_0, \psi$ are the modular parameters of the solution, 
$\cot\psi=\lambda$. It is clear that these
solitons lie in the complement of the maximal torus since $U$ and $W$ 
do not vanish for any value of $x$
unless $x$ goes to infinity. The topology of the moduli space is
$S^1\times \bR$: The modular parameter $x_0$ is due to the
translational invariance of the
underlying theory while the modular angular parameter $\psi$ is due 
to the charge $Q$ given in \fourbb.  These
are reminiscent to the modular parameters of BPS monopoles in
four-dimensions (see for example [\athit]).

However, there is a more general class of $SU(2)$ solutions for which 
$N\not=\kappa$.  To find these new
solitons, we use the angular parameterisation and the equations 
\ninedaan\  for $n=1$ become
$$
\eqalign{
{d\over dx}\theta&=\pm m[-\kappa \cos \phi+\sin(\theta)]
\cr
{d\over dx}\phi&=\pm m\kappa \cot\theta \sin\phi\ .}
\eqn\ninedaa
$$
As it can been seen from
\vacthree, the $SU(2)$ $n=1$ model has two vacua $(\theta_0, 0)$ and 
$(\pi-\theta_0, 0)$. We remark that a  
linearisation of \ninedaa\ at these two vacua reveals that one of them
is a source
while the other is a sink.  The differential equations \ninedaa\ is a 
Hamiltonian flow for the function
$$
\alpha(\theta, \phi)=-{1\over \sin\phi \sin\theta}+{1\over \kappa} \cot\phi
\eqn\hamone
$$
with symplectic form
$$
\Omega=\pm  {1\over (m \,\kappa )\, \sin^2\phi \sin\theta}\, 
d\theta\wedge d\phi \ .
\eqn\hamtwo
$$
Since $\alpha$ is preserved by the flow, we can rewrite \hamone\ as
$$
\sin\theta={\kappa\over \cos\phi-\alpha \kappa \sin\phi}\ ,
\eqn\moned
$$
where $\alpha$ is a real constant.
The solutions of \ninedaa\ in the $x$ parameterisation are
$$
\eqalign{
\cot\phi\equiv z&={\kappa\over 2 \sqrt{1-\kappa^2}}\exp{\big[\pm m 
\sqrt{1-\kappa^2} (x-x_0)\big]}
\cr & \qquad\qquad
\bigg([\exp{\big[\mp m \sqrt{1-\kappa^2} (x-x_0)\big]}+{\alpha\over 
\sqrt{1-\kappa^2}}]^2+1-\alpha^2\bigg)
\cr
\cos\theta&=\cases{{\sqrt{(z-\alpha\kappa)^2-\kappa^2(1+z^2)}\over 
z-\alpha\kappa}\ , \qquad -\infty< x
<x_{min}
\cr 
-{\sqrt{(z-\alpha\kappa)^2-\kappa^2(1+z^2)}\over z-\alpha\kappa}\ , 
\qquad x_{min}< x <+\infty\ ,}}
\eqn\mtwod
$$
where at $x=x_{min}$, $z(x)$ takes its absolute minimum value.  To 
verify that \mtwod\ solves the equations
\ninedaa, we differentiate $z$ with respect to $x$ and use the
equation for $\phi$ to
eliminate the derivative of $\phi$ from the expression.  The $x$ 
dependence can also be eliminated by inverting 
$z$ to express the exponential of $x$ in terms of $z$. However this 
equation is quadratic and there are two
possible solutions distinguished by a sign. To satisfy the equations
\ninedaa, one has to choose for
$-\infty< x <x_{min}$ the solution with the plus sign and for 
$x_{min}< x <+\infty$ the solution with the
minus sign together with the corresponding expression for 
$\cos\theta$ in \mtwod. 

To find the solitons of the $SU(2)$ $n=1$ model, we substitute 
\mtwod\ in \twob\ and observe that they
are {\it time-dependent}.   For $\alpha=0$, \mtwod\ reduces to 
the solution
\eightd\ given above. Moreover, after the redefinition  
$x_0\rightarrow x_0\pm {1\over
m\sqrt{1-\kappa^2}}\rm{log}(|\alpha| {1-\sqrt{1-\kappa^2}
\over\sqrt{1-\kappa^2}})$, \mtwod\ reduces in the
limit $\alpha\rightarrow \pm \infty$ to the embedded SO(2) 
solitons found in the previous section.
The charge and the mass of the solutions \mtwod\ are
$$
Q=\mp 4 \sqrt{1-\kappa^2} e^\sigma \ ,
\eqn\mthreed
$$
and
$$
E= 2 m \sqrt{1-\kappa^2} e^\sigma\ ,
\eqn\mfourd
$$
respectively,  where
$$
e^\sigma=1+{\alpha \kappa^2\over \sqrt{1-\kappa^2} \sqrt{1+\alpha^2 
\kappa^2}}\bigg[- {\pi\over2}+{\rm
arc}\tan \big({\alpha \kappa^2\over \sqrt{1-\kappa^2} \sqrt{1+\alpha^2
\kappa^2}}\big)\bigg]\ .
\eqn\msixd
$$
The constant $\alpha$ is not a  modular parameter because the mass 
and the charge 
depend upon it. For example, the mass of the $\alpha=0$ solutions 
\eightd\ is \foot{This corrects
the expression for the mass of these solutions in [\gpta].}
$$
E=2 m \sqrt{1-\kappa^2}\ .
\eqn\msixdd
$$
Note that the mass \mfourd\ of the solitons as a function of $\alpha$ 
has critical points at
$\alpha\rightarrow\pm\infty$. It turns out that $E(+\infty)$ is the 
absolute minimum
and is the value of the mass of the embedded static solution \fivec\ 
for $n=1$, while
$E(-\infty)$ is the absolute maximum
and is the value of the mass of the embedded static solution \fivecc\ 
for $n=1$. This is in agreement
with the corresponding limits of the solution \mtwod\ mentioned above.
The moduli space of the solutions
\mtwod\ is again a cylinder with co-ordinates
 $(x_0, \psi)$ and with metric  
$$
ds^2=e^\sigma {\sqrt{1-\kappa^2}\over m
(1+\alpha^2 \kappa^2)}\bigg[m^2 
(1-\kappa^2+\alpha^2\kappa^2)  dx_0^2+ d\psi^2\bigg]\ .
\eqn\mfived
$$
We remark that there is a  transformation of the moduli co-ordinates 
$(x_0, \psi)\rightarrow (y, \chi)$ such
that the moduli space metric above can be written in the form 
$ds^2={E\over2}(dy^2+d\chi^2)$.


\chapter{The SU(2) n=2 model}

To show that all solitons of the $SU(2)$ $n=2$ model lie on the maximal
torus of $SU(2)$,  we begin from the  equations
\fivedn\ for
$M,N$ with $n=2$ 
$$
\eqalign{
{d\over dx}M\pm m [-\kappa N+2(1-M^2)M]&=0
\cr
{d\over dx}N\pm m[\kappa M-2NM^2]&=0 \ .}
\eqn\fived
$$
 We then use the field redefinitions
$$
X={N\over M}, \qquad Y={N\over \sqrt {1-M^2}}
\eqn\tend
$$
to simplify \fived\ to
$$
\eqalign{
{d\over dx}X\pm  m\big[ \kappa (1+X^2)-2X\big]&=0
\cr
{d\over dx}Y\pm  m \kappa X^{-1} Y [1-Y^2]&=0\ .}
\eqn\elevend
$$
The solutions in terms of $X,Y$ for $0<\kappa<1$ are
$$
\eqalign{
X&={1\over \kappa}\bigg(1+\sqrt{1-\kappa^2} \tanh\big(\pm m 
\sqrt{1-\kappa^2}(x-x_0)\big) \bigg)
\cr
{Y^2\over 1-Y^2}&={\rho}^{-1} X^2 \exp\big(\mp 2m (x-x_0)\big)\cosh^2\big(
m\sqrt{1-\kappa^2} (x-x_0)\big)\ ,}
\eqn\twelved
$$
and
$$
\eqalign{
X&={1\over \kappa}\bigg(1+\sqrt{1-\kappa^2} \coth \big(\pm m 
\sqrt{1-\kappa^2} (x-x_0)\big) \bigg)
\cr
{Y^2\over 1-Y^2}&={\rho}^{-1} X^2 \exp\big(\mp 2m (x-x_0)\big) \sinh^2\big(
m\sqrt{1-\kappa^2} (x-x_0)\big)\ ,}
\eqn\twelved
$$
where $\rho, x_0$ are real constants ($\rho\geq 0$).  It remains 
to examine whether or not these solutions
interpolate between the different vacua of the theory. As we have 
already mentioned in section four, the
$SU(2)$ $n=2$ model has four vacua. It turns out that the 
solutions \twelved\ interpolate between two different vacua only for
$\rho=0$, in which case
$U=W=0$. This implies that all solitons of the $SU(2)$ $n=2$ model 
lie on the maximal torus.  The
same applies for all solitons of the $SU(2)$ $n=2$  model with $\kappa=1$.

To explain why there are no solitons of the $SU(2)$ $n=2$ model
that lie in the complement of the maximal torus of $SU(2)$, let 
us study the vacuum structure of this model
in more detail.  A linearisation of the equations \ninedaan\ around 
the vacua \vacthree\  reveals that 
$({\theta_0\over2}, 0)$ and $({-\theta_0+2\pi \over2}, \pi)$ are 
sources, and $({\theta_0+\pi
\over2},\pi)$ and $({-\theta_0+ \pi \over2}, 0)$ are saddles (for the
Bogomol'nyi equations with the plus sign).  Solving the  equations  
\ninedaan\ with $n=2$
for $\theta$ and $\phi$ by setting $\phi=0,\pi$, we find precisely 
the solitons of the
$SU(2)$ $n=2$ model that lie on the maximal torus of $SU(2)$ and 
therefore are
embeddings of the $SO(2)$ solitons.  These solitons interpolate 
between the source ($\phi=0$) and the saddle
($\phi=0$), the saddle ($\phi=0$) and the source ($\phi=\pi$), 
the source ($\phi=\pi$) and the saddle
($\phi=\pi$), and the saddle ($\phi=\pi$) and the source 
($\phi=0$). Any additional solutions should
interpolate either between the two saddles or between the 
saddles and the sources.  But the only directions
from the two saddles which are not connected to another vacuum 
are those that point outwards. Therefore no
soliton can exist that starts from one saddle to go to the other 
or from a saddle to go to a source.  In fact,
the outward directions from the saddles and some of the directions 
from the sources are connected to a sink at
$$
(\theta, \phi)=({\pi\over 2}, {\pi\over2})\ ,
\eqn\fournew
$$
which is a fixed point of the flow described by the equations 
\ninedaan, $n=2$, but
{\it not} a vacuum of the theory.  We shall conjecture in the 
next section that a similar behaviour occurs in
all $SU(2)$ models with $n$
even.


\chapter{The SU(2) $n>2$ models}

The equations \fivedn\ or \ninedaan\ for $SU(2)$ models with $n>2$ 
and $\kappa\not= 0$ are
rather involved and we have not been able to find the solitons 
that lie in the complement of the maximal
torus of $SU(2)$. Another way to proceed is to investigate the 
qualitative properties of the solutions
of  these equations by linearising them about the vacua of the 
theory. This will allow us to find the pairs of vacua
that are connected by solitons. The results of this analysis for 
$n=3,4,5$ have being confirmed with a
numerical computation (see Figs. [1,2,3]). 

It is sufficient to examine the linearisation properties of the 
equations \ninedaan\ in 
the interval $0\leq
\theta <{\pi\over2}$.  This is because the equations \ninedaan\ 
are invariant under the
discrete symmetries
$$
(\theta, \phi, x)\rightarrow (\pi-\theta, \pi-\phi, x)
\eqn\nthreed
$$
for $n$ even, and
$$
(\theta, \phi, x)\rightarrow (\pi-\theta, \phi, -x)
\eqn\nfourd
$$
for   $n$ odd, which can be used to extend the analysis to 
the whole range of $\theta$, ($\theta\in [0,\pi)$).
Another property of the equations \ninedaan\ for $n$ {\it even} 
is that, apart from the vacua of the
theory, there is another fixed point at
$(\theta, \phi)=({\pi\over2}, {\pi\over2})$.  

Next we order the vacua $(\theta,\phi)$ in \vacthree\ with 
respect to the value of
$\theta$, and denote these values by $\theta_i$, $i=1,\dots, 
2n$, {\rm i.e.} $\theta_1<\dots
\theta_i<\theta_{i+1}\dots<\theta_{2n}$. For $\theta$ in 
$[0,\pi/2)$, a linearisation of
\ninedaan\ with the plus sign reveals that the vacua $\theta_{4q+1}$ are
sources, the vacua
$\theta_{4q+3}$ are sinks and the rest are saddle points. 
Observe that the
sources $\theta_{4q+1}$ and the saddles $\theta_{4q+2}$  
lie at the
$\phi=0$ semicircle, while the sinks $\theta_{4q+3}$ and the 
saddles $\theta_{4q+4}$  lie at the $\phi=\pi$
semicircle.   First, we shall consider the
solutions interpolating amongst vacua $\theta_i$ in the interval 
$[0,\pi/2)$.  It is
expected that there should be a one parameter family of solitons 
that interpolate between the source
$\theta_{4q+1}$  and the sink $\theta_{4q+3}$, and the same source  
and the sink $\theta_{4q-1}$. There
also should be a unique soliton interpolating between the source 
$\theta_{4q+1}$ and the saddle $\theta_{4q}$,
and the sink $\theta_{4q+3}$ and the saddle
$\theta_{4q+2}$. In addition to these solutions, we have the 
solitons that lie on the maximal torus.
$$\vbox{
    \beginlabels\refpos 71 481 {}
            \put 190 423 {\nwarrow}
            \put 326 423 {\nwarrow}
            \put 215 354 {\downarrow}
            \put 344 354 {\downarrow}
            \put 190 286 {\nearrow}
            \put 326 286 {\nearrow}
            \put 134 262 {\theta_1}
            \put 134 309 {\theta_2}
            \put 134 401 {\theta_5}
            \put 134 447 {\theta_6}
            \put 419 331 {\theta_3}
            \put 419 377 {\theta_4}
	    \put 130 462 {\pi}
            \put 124 355 {\pi/2}
            \put 146 239 {0}
	    \put 408 239 {\pi} 
            \put 271 235 {\pi/2} 
            \put  99 348 {\uparrow}
            \put 268 218 {\rightarrow}
            \put 100 364 {\theta}
            \put 284 216 {\phi}
            \endlabels
            \epsfxsize=.65\hsize
            \centerline{\epsfbox{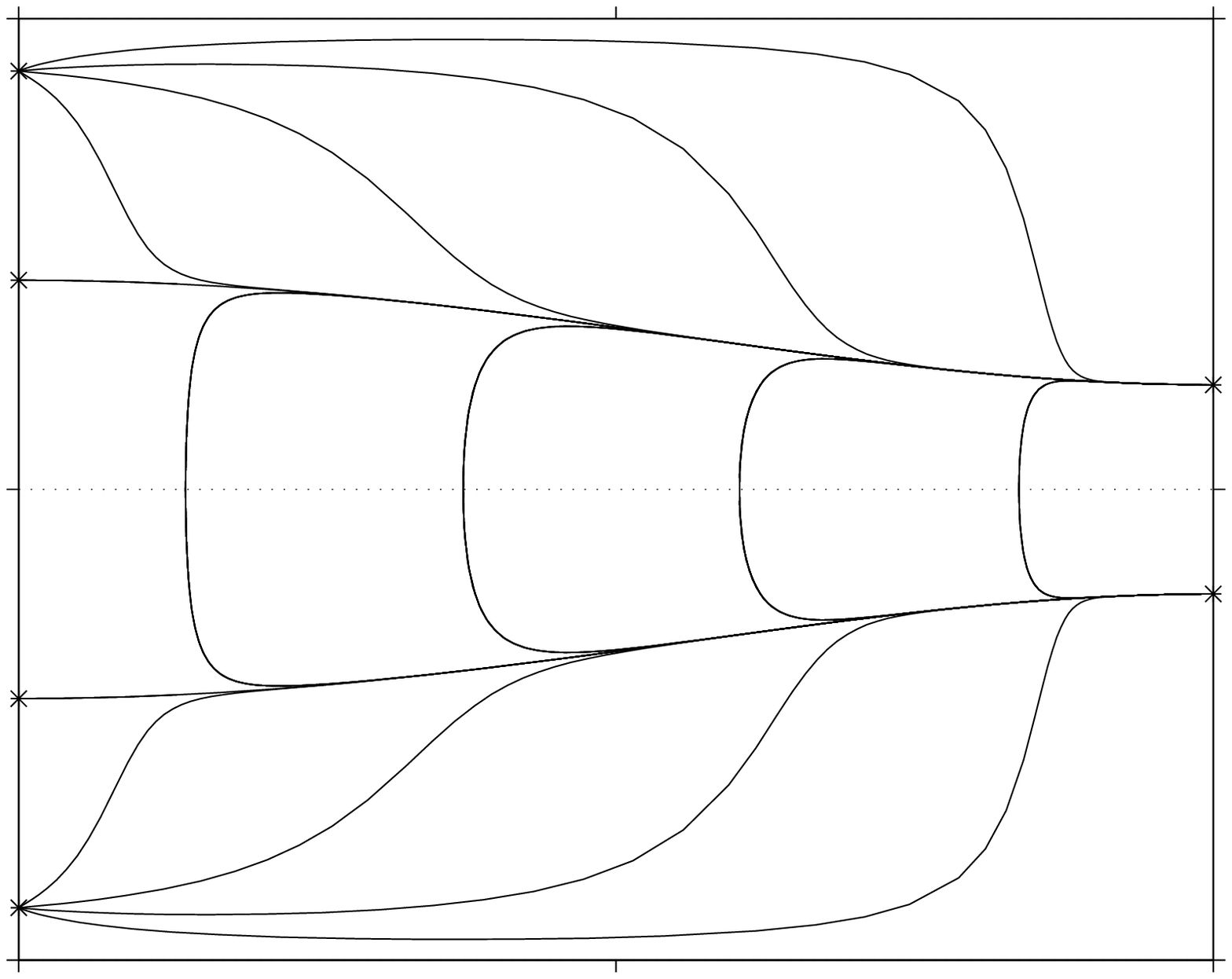}}}
$$
\nobreak\vskip-\medskipamount
\centerline{{\bf Fig.~1:} The solutions of the ${\rm SU(2)}$ $n=3$ model}

The solitons interpolating amongst the vacua in $[\pi/2, \pi)$ are
similar to those for the vacua in $[0,
\pi/2)$ examined above.  This follows immediately from the discrete 
symmetries \nthreed\ and \nfourd\ of
the equations \ninedaan. These symmetries act on $(\theta, \phi)$ 
that lie on the disc constructed from the square
$[0,\pi]\times[0,\pi]$ by identifying the points $(0,0)$ and
$(0,\pi)$, and the points $(\pi,0)$ and
$(\pi,\pi)$. For 
$n$ even,  the discrete symmetry is just the antipodal map of the disc leaving
the centre
$(\pi/2,\pi/2)$ fixed, whereas for 
$n$ odd, it is a reflection at the hyperplane $\theta=\pi/2$ together 
with a reversal of the direction of the
flow and an interchange sinks and sources.
$$\vbox{
    \beginlabels\refpos 71 725 {}
            \put 219 690 {\searrow}
            \put 223 649 {\swarrow}
            \put 221 610 {\uparrow}
            \put 223 567 {\searrow}
            \put 219 528 {\nearrow}
            \put 326 690 {\searrow}
            \put 331 649 {\swarrow}
            \put 328 610 {\uparrow}
            \put 331 567 {\searrow}
            \put 326 528 {\nearrow}
            \put 132 512 {\theta_1}
            \put 132 540 {\theta_2}
            \put 419 554 {\theta_3}
            \put 419 581 {\theta_4}
            \put 132 595 {\theta_5}
            \put 132 623 {\theta_6}
            \put 419 637 {\theta_7}
            \put 419 664 {\theta_8}
            \put 132 678 {\theta_9}
            \put 132 705 {\theta_{10}}
	    \put 128 716 {\pi}
            \put 122 609 {\pi/2}
	    \put 146 493 {0}
	    \put 408 493 {\pi} 
            \put 271 489 {\pi/2} 
            \put  99 602 {\uparrow}
            \put 268 472 {\rightarrow}
            \put 100 618 {\theta}
            \put 284 470 {\phi}
            \endlabels
            \epsfxsize=.65\hsize
            \centerline{\epsfbox{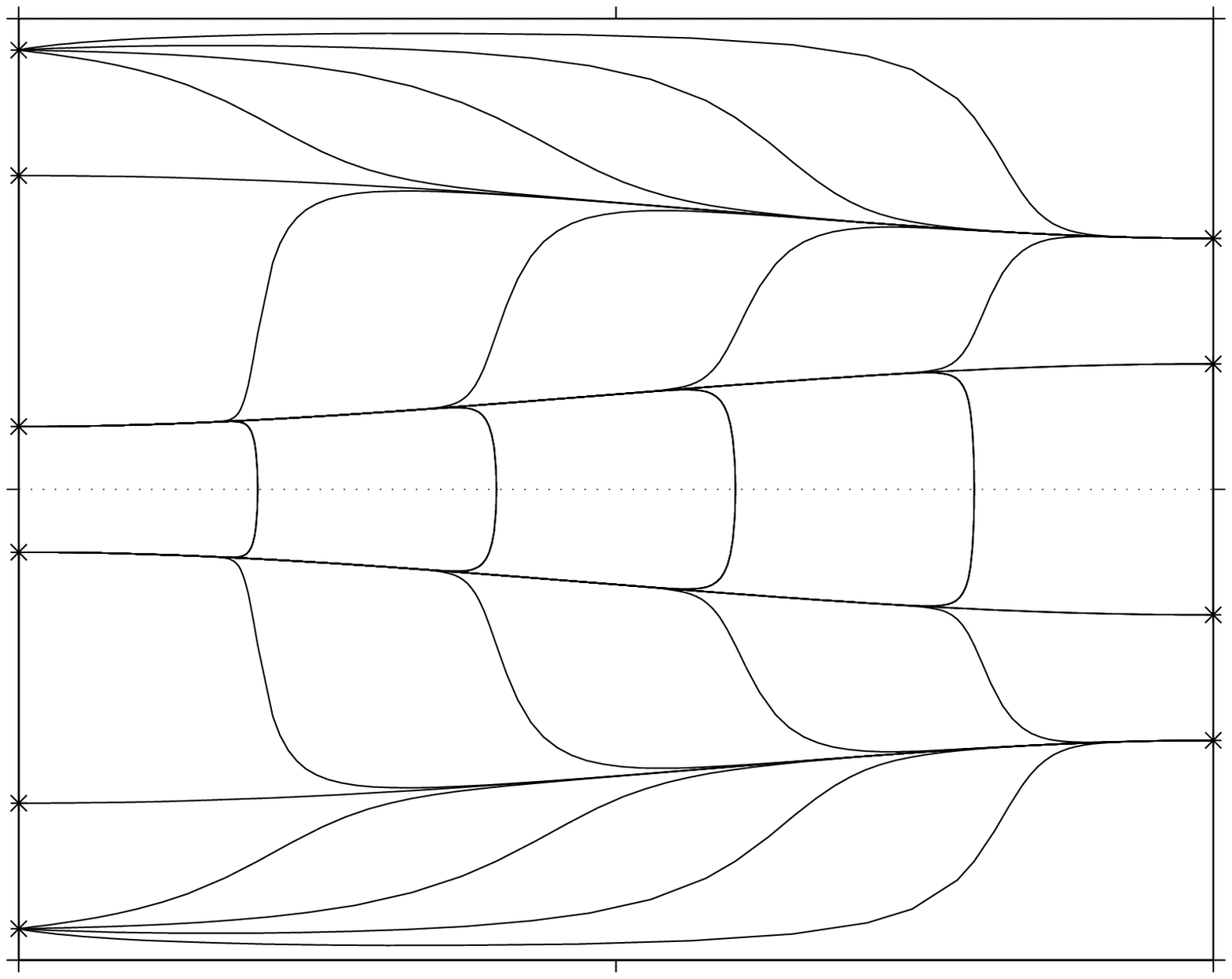}}}
$$
\nobreak\vskip-\medskipamount
\centerline{{\bf Fig.~2:} The solutions of the ${\rm SU(2)}$ $n=5$ model}

Let us consider the solutions amongst the vacua that lie near the 
boundary of $[0,\pi]$. The lowest
vacuum $\theta_1$ (which is a source) connects only to the sink 
$\theta_3$ through solitons that lie in the
complement of the maximal torus of $SU(2)$. Similarly the highest 
vacuum $\theta_{2n}$ (which is a sink)
connects only to the source $\theta_{2n-2}$. It remains to examine 
the solutions interpolating amongst
vacua that lie `near' $\theta=\pi/2$. For $n=4\ell+1$, $\ell\in \bN$, 
the highest vacuum in $[0,\pi/2)$ is the
source $\theta_n$ and this connects to the sink $\theta_{n-2}$ and to 
the saddle $\theta_{n-1}$ as it has been
described above for the vacua that lie within $[0,\pi/2)$.  In
addition, 
$\theta_n$  connects with the lowest
vacuum in $[\pi/2,\pi)$ which is the sink $\theta_{n+1}$ at the
$\phi=0$ 
semicircle; the solutions are similar
to the ones that we have found in the $SU(2)$ $n=1$ model in section
five. The sink $\theta_{n+1}$ also
connects to the source
$\theta_{n+3}$ and the saddle $\theta_{n+2}$ as required by the 
reflection symmetry.  It is straightforward to
repeat this analysis for the $n=4\ell+3$ case. In figures 1 and 2, 
we confirm our results by a numerical
computation for the solutions of equation \ninedaan\  of the
$SU(2)$ $n=3, 5$ models that lie in the complement of the maximal
torus of $SU(2)$.
We remark that it is possible to  interpolate amongst the vacua of 
$SU(2)$ model with $n\geq 5$ in
a way different from the one proposed above.  However, the numerical 
calculation for $n=5$ supports the case
that has been presented.
$$\vbox{
    \beginlabels\refpos 71 698 {}
            \put 220 648 {\swarrow}
            \put 220 599 {\nwarrow}
            \put 220 563 {\swarrow}
            \put 220 535 {\searrow}
            \put 220 498 {\nearrow}
            \put 326 649 {\swarrow}
            \put 326 616 {\nwarrow}
            \put 326 586 {\nearrow}
            \put 326 550 {\searrow}
            \put 326 498 {\nearrow}
            \put 265 586 {\leftarrow}
            \put 283 569 {\rightarrow}
            \put 134 479 {\theta_1}
            \put 134 514 {\theta_2}
            \put 419 531 {\theta_3}
            \put 419 565 {\theta_4}
            \put 131 588 {\theta_5}
            \put 134 618 {\theta_6}
            \put 419 635 {\theta_7}
            \put 419 669 {\theta_8}
            \put 275 576 {\times}
	    \put 130 682 {\pi}
            \put 122 575 {\pi/2}
            \put 146 459 {0}
	    \put 408 459 {\pi} 
            \put 271 455 {\pi/2} 
            \put  99 568 {\uparrow}
            \put 268 438 {\rightarrow}
            \put 100 584 {\theta}
            \put 284 436 {\phi}
            \endlabels
            \epsfxsize=.65\hsize
            \centerline{\epsfbox{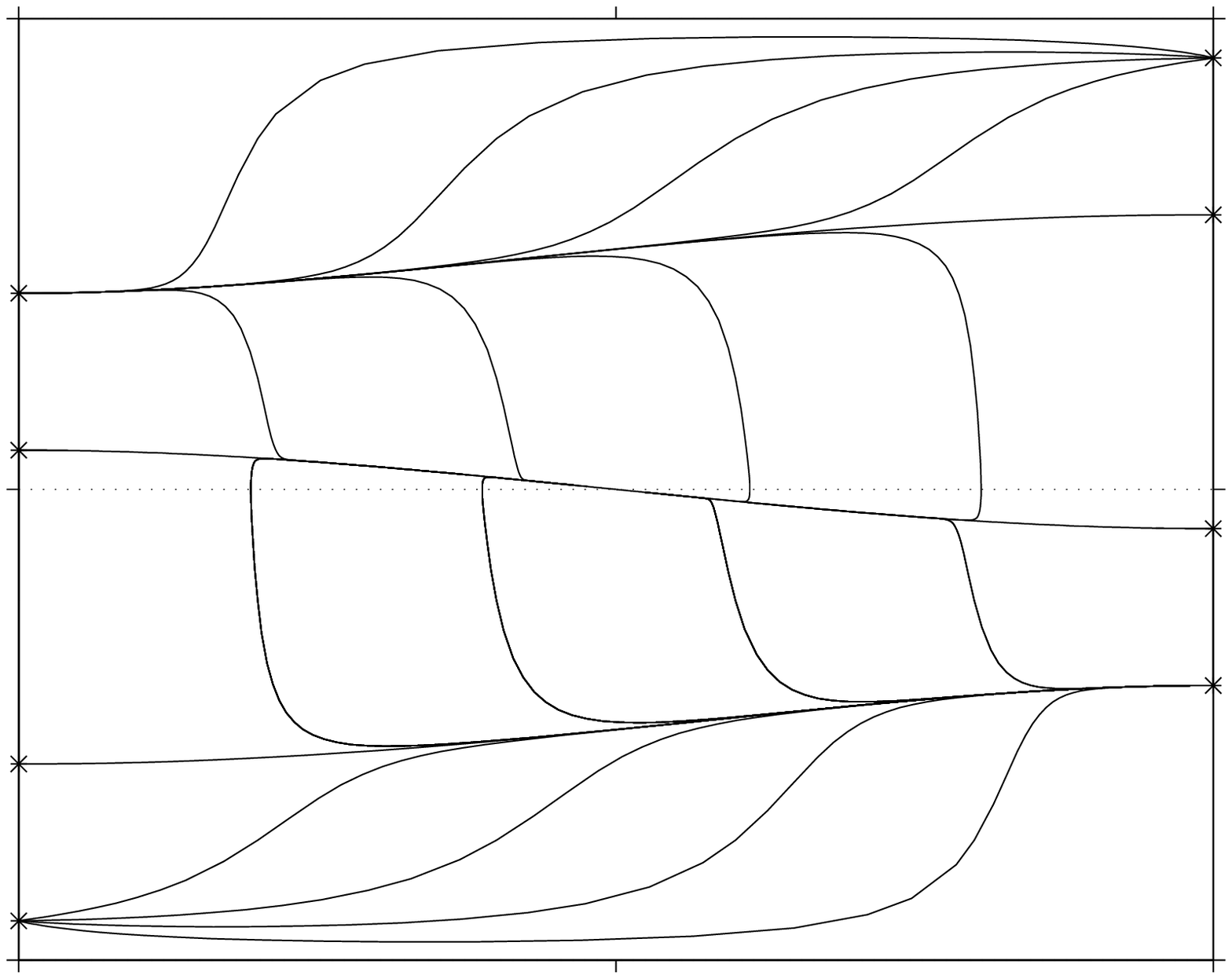}}}
$$
\nobreak\vskip-\medskipamount
\centerline{{\bf Fig. 3~:} The solutions of the ${\rm SU(2)}$ $n=4$ model}

Next  we turn to examine the  $n$ even
case and observe that  $({\pi\over2}, {\pi\over2})$ is a fixed point
of \ninedaan\ but {\it not} a vacuum of
the theory;
$({\pi\over2}, {\pi\over2})$  is  source for $n=4 \ell$ and a sink for
$n=4\ell+2$. 
For $n=4\ell+2$, the sink $\theta_{n-1}$ connects to the saddle 
$\theta_{n-2}$  and to the
source $\theta_{n-3}$ in the usual way. Both the sink $\theta_{n-1}$ 
and the saddle $\theta_n$ connect to   
the $({\pi\over2}, {\pi\over2})$  fixed point.   Using the antipodal 
map, we can easily
determine the behaviour of the solutions associated to the vacua
$\theta_{n+1}$ and $\theta_{n+2}$ from the solutions associated to 
the vacua $\theta_n$ and $\theta_{n-1}$. We
remark that for
$n=2$, we recover the behaviour we have found in section six.  It 
is straightforward to repeat
this analysis for the $n=4\ell$ case. In figure 3, we confirm our 
results with a numerical
computation for the solutions of \ninedaan\  of the
$SU(2)$ $n=4$ model that lie in the complement of the maximal torus of
$SU(2)$.

\chapter{The $G$-models}

In the previous sections we have investigated the soliton solutions of
sigma models with target
space the groups $SO(2)$ and $SU(2)$.  In this section, we shall 
explore some of the properties of the soliton
solutions of $G$-models for $G$ any semisimple compact Lie group. 
The vacua of any $G$-model can be arranged to lie on a maximal torus 
of the group $G$.
To show this, we first choose, without loss of generality, $\kappa$ in
a Cartan subalgebra of $G$ as in section two, i.e.
$$
\kappa=\kappa^m t_m\ ,
\eqn\gone
$$ 
where $t_m$ is an orthonormal basis of the Cartan subalgebra and 
$\kappa^m$ are the components of $\kappa$
 ($\kappa^m\in \bR$). 
Now suppose that there is a vacuum $k_0$ that does not lie on this 
maximal torus of $G$, then there is a
$h_0\in G$ such that  $k'_0=h_0 k_0 h_0^{-1}$ and $k_0'$ is in the 
maximal torus. Using  the invariance of
the equation for the vacua under conjugation, \fourb\ can be written as
$$
2h^{-1}_0\kappa h_0+\sum _A t_A{\rm tr}\big(t_A(k'_0)^n\big)=0\ ,
\eqn\goneaa
$$
where $\kappa$ is as in \gone.
Now, this equation has solutions if and only if $h_0\kappa h_0^{-1}$ 
is again in the Cartan
subalgebra of $G$.  It follows then that $h_0$ is an element of the 
Weyl group of $G$ in which case
$k_0$ is in the maximal torus. Therefore all vacua of the $G$-models 
are on the maximal torus of $G$.
This allows us to set 
$$
k=\exp\big({\theta^m t_m}\big)
\eqn\gtwo
$$
and rewrite the equation for the vacua as
$$
2\kappa_m+{\rm tr}\big(t_m \exp({n\theta^r t_r})\big)=0\ .
\eqn\gthree
$$    
Observe that the second term in the equation for the vacua is the 
derivative of the character
$\chi(\theta)={\rm tr}\, k$ of $G$ with respect to the co-ordinates 
of the maximal torus.

As in the $SU(2)$ model, the $G$-model also has two classes of solitons. 
(i) The static
 solitons that lie on the maximal torus of $G$. 
(ii) The time-dependent solitons, apart from their asymptotic values, 
lie in the complement of the maximal
torus of $G$.  In the former case, the equations
\threeb\  can be written as
$$
{d \over dx}\theta^m(x)=\mp {m\over2} \bigg(2\kappa^m + {\rm tr}
\big(t^m \exp(n\theta^r t_r)\big)\bigg)\ .
\eqn\gfour
$$
A class of solutions of these  equations can be found by setting  
all but one angle, say $\theta^1$, to their
vacuum values.  The solitons obtained in this way  are precisely 
those that are embeddings of the
$SO(2)$ solitons along the direction of $\theta^1$. There are as many 
independent directions for
embedding
$SO(2)$ solitons in the $G$-model as the rank of the group $G$.  Apart
from these solitons, there are other
static solitons particular to the $G$-model ($\rm{rank}\,{\cal
L}(G)>1$) for which more than one of  the
angles
$\theta^m$ are not set to their vacuum values, and the equations 
\gfour\ do not separate to a sum
of $SO(2)$ ones.   To see this, let us consider the
$SU(3)$ model.  The maximal torus in this case is $T^2$ and the theory
has four vacua for some choice of $\kappa$. The  equations \gfour\ are
similar to the standard Morse flow of the height function $f$ of
$T^2$.  Apart from the Morse flows of $f$ which one gets by embedding 
the Morse flows of $S^1$ along the
two homology cycles of $T^2$ and which are similar to the embedding of
$SO(2)$ solitons in the $SU(3)$ model,
one obtains additional flows, for example an one-parameter family of 
flows from the absolute maximum to the
absolute minimum of $f$.

The investigation of the time-dependent solitons that lie in the 
complement of the maximal torus of $G$ is
more involved because, apart from the Bogomol'nyi equations along 
the maximal torus, we should also solve the
Bogomol'nyi equations along the root directions of $G$.  Despite 
this, some of them can be
obtained from embedding the $SU(2)$ time-dependent solitons into the
$G$-model up to a sigma model transformation \fiveb\ of the $G$ model.
This is done by embedding the $SU(2)$
group into the $G$ group up to a conjugation. This group theoretical 
problem has been investigated at the Lie
algebra level in [\Dynkin] and it was found that there is at least 
one embedding of $SU(2)$ in $G$ for every
positive root $\alpha$ of $G$. For such an embedding the coupling 
constant $\kappa$ of the $G$-model is
$$
\kappa^m=\tilde\kappa \alpha^m \ ,
\eqn\gtwo
$$
where $\tilde\kappa$ is a real number and $|\tilde \kappa|
<1/|\alpha|^2$, {\rm i.e.} $\kappa$ is along a
direction in the Cartan subalgebra of
$G$ labeled by the root $\alpha$. The generators of the
$SU(2)$ subgroup of $G$ are
$$
i  \tilde h_\alpha, \qquad i(\tilde E_\alpha+\tilde E_{-\alpha})\ , 
\qquad \tilde E_\alpha-\tilde E_{-\alpha} \ ,
\eqn\gthree
$$
where $\{\tilde h_\alpha ,\tilde E_\alpha, \tilde E_{-\alpha}\}$ 
are the generators of ${\cal L}(G)$ in the
Chevalley basis. 
It is clear from \gtwo\ that all these embeddings do not describe 
the most general soliton solutions of the
$G$-model.  For this, one should solve the Bogomol'nyi equations 
of the $G$-model for $\kappa$ along a
generic direction in the Cartan subalgebra of $G$.

We expect that apart from the one-soliton solutions that we have 
discussed so far, the $G$-models may
have  multi-soliton solutions as well.  Indeed for $G$-models 
with $\kappa=0$, we can simply
embed the multi-soliton solutions of the sine-Gordon theory 
(see for example [\raj] and references within.).
However, since the multi-soliton solutions of the sine-Gordon 
theory are not solutions of the Bogomol'nyi
equations, the same applies for embedded multi-soliton solutions 
in the $G$-model.

\chapter{Concluding remarks and summary}

Quantum mechanically, some solitons may decay to other
solitons that carry the same topological charge but different Noether
charge by emitting radiation.  This is because, for $G$-models 
for which $G$ is a semisimple
group, the fundamental states of the
theory  as well as its solitons are charged with respect to the 
{\it same} Noether charge $Q_N$. 
Note that the solitons, in addition to the Noether charge $Q_N$, 
carry topological charge $Q_T$.  From
charge conservation it follows that a soliton with topological 
charge $Q_T$ and with  Noether charge
$Q_N$ may decay to an other soliton with the same topological charge but 
with Noether charge $Q_N'$ and
$|Q_N'|<|Q_N|$, and some fundamental states with Noether charge $Q_N-Q'_N$. 
Energy conservation though seems to rule out such a process because
the fundamental states do not saturate the bound
($\kappa\not=1$). However if for some unknown mechanism to us such a
process is allowed the stable soliton configurations are those that
have the least mass for given topological charge. 
An example of such a configuration is the soliton \fivec\ of the $SU(2)$
$n=1$ model. Moreover, as it is well known, the coupling
constant of $G$-models, $G$ semisimple, is quantised due to the 
presence of the torsion term and therefore it
may be more appropriate to develop a non-perturbative method, 
similar to that developed for the
Wess-Zumino-Witten model [\witten], to investigate the fundamental 
states of the theory instead of linearising
the theory about a vacuum in the weak coupling limit.  Another 
related issue is the quantisation of the charge
$Q$ of the solitons.  Since
$Q$ is not purely topological, the quantisation of $Q$ does not 
follow from classical considerations;  a
quantisation of the moduli space suggests though that $Q$ should 
be quantised provided that the moduli
coordinate $\psi$ is periodic which is the case whenever the orbits 
of the vector field $X$ in $G$ are periodic. We also remark that even
if $Q$ is quantised the
usual stability argument for BPS monopoles does not apply to this case
because the bound for the energy is in terms of one charge rather than
two, which are necessary to establish the stability for the solitons
using the triangular inequality.

To summarise, we have investigated the soliton solutions of 
(1,1)-super\-symmetric massive sigma models with
torsion and target space a  group $G$ for a class of scalar 
potentials characterised by a coupling constant
$\kappa$ and an integer $n$. These solitons are solutions of 
Bogomol'nyi equations which arise from the
saturation of a bound for the energy of these models in terms of 
a charge $Q$ that appears as a central charge
in the (1,1)-supersymmetry algebra. The charge $Q$ is the sum of 
a Noether charge $Q_N$ and a topological
charge $Q_T$. The $G=SO(2)$ model is the simplest of the $G$-models 
and its solitons can be
easily computed; the
$SO(2)$ model with $\kappa=0$ is the supersymmetric sine-Gordon 
theory.  For $G$ a semisimple group there are two
classes of solitons to consider, one is a set of static solitons 
that lie on a maximal torus of
$G$ and the other is a set of time-dependent solitons that, apart 
from their asymptotic values, lie in the
complement of a maximal torus of
$G$. For $G=SU(2)$, we have found all static solitons as embeddings 
of the corresponding $SO(2)$ solitons. In
addition,  we have explicitly computed the time-dependent solitons 
of the $n=1$ model, and we have
shown that all solitons of the $n=2$ model are static. We have also 
presented
the qualitative properties of the time-dependent solitons of the 
$SU(2)$ $n>2$ models and confirmed our
results with a numerical calculation for the
$n=3,4,5$ ones.  For sigma models with target space a semisimple 
group $G$, some of their solitons can
be obtained from embedding the solitons of the $SO(2)$ model and 
the solitons of the $SU(2)$ model with the
corresponding value of $n$.

\vskip 0.5cm

\noindent{\bf Acknowledgments:}  G.P. is supported by a 
University Research Fellowship from the Royal Society. We would 
like to thank A. Sornborger and P.K. Townsend
for helpful comments.

\refout

\bye